\documentclass[%
 reprint,
superscriptaddress,
 amsmath,amssymb,
 aps,
floatfix,
]{revtex4-2}
\usepackage{comment}
\usepackage{tabularx}
\usepackage{graphicx}% Include figure files
\usepackage{dcolumn}% Align table columns on decimal point
\usepackage[normalem]{ulem}
\usepackage{bm}% bold math
\usepackage{siunitx}
\begin{document}
 
\preprint{APS/123-QED}
\title{Quantum Advancements in Neutron Scattering Reshape Spintronic Devices}

\author{M. E. Henderson}
\email{hendersonme@ornl.gov}%
\affiliation{Neutron Technologies Division, Oak Ridge National Laboratory, Oak Ridge, TN 37831, USA}
\affiliation{Institute for Quantum Computing, University of Waterloo, Waterloo, ON, Canada, N2L3G1}
\affiliation{Department of Physics \& Astronomy, University of Waterloo,
  Waterloo, ON, Canada, N2L3G1}

\author{D. G. Cory}
\affiliation{Institute for Quantum Computing, University of Waterloo, Waterloo, ON, Canada, N2L3G1}
\affiliation{Department of Chemistry, University of Waterloo, Waterloo, ON, Canada, N2L3G1}
\date{\today}

\author{D. Sarenac}
\affiliation{Institute for Quantum Computing, University of Waterloo, Waterloo, ON, Canada, N2L3G1}
\affiliation{Department of Physics, University at Buffalo, State University of New York, Buffalo, New York 14260, USA}

\author{D. A. Pushin}
\email{dmitry.pushin@uwaterloo.ca}
\affiliation{Institute for Quantum Computing, University of Waterloo, Waterloo, ON, Canada, N2L3G1}
\affiliation{Department of Physics \& Astronomy, University of Waterloo,
  Waterloo, ON, Canada, N2L3G1}

\begin{abstract}
Topological magnetism has sparked an unprecedented age in quantum technologies. Marked by twisted spin structures with exotic dynamical modes, topological magnets have motivated a new generation of spintronic devices which transcend the limits of conventional semiconductor-based electronics. While existing material probes have biased studies and device conceptualizations for thin samples in two dimensions, advancements in three-dimensional probing techniques using beams of neutrons, are transforming our understanding of topological and emergent physics to reimagine spintronic devices. Here, we review recent neutron scattering breakthroughs which harness quantum degrees of freedom to enable three-dimensional topological investigations of quantum materials. We discuss applications of structured and tomographic neutron scattering techniques to topological magnets, with particular emphasis on magnetic skyrmion systems and their inspired three-dimensional logic device infrastructures through novel multi-bit encoding and control schemes. SANS-based dynamic visualizations and coherent manipulations of three-dimensional topological qubits are proposed using electric field controls of depth-dependant helicities and spin-orbit tuning of the neutron beam. Together, these investigations uncover a new world of three-dimensional topological physics which enhances spintronic devices through a novel set of structures, dynamics, and controls, unique to three-dimensional systems.

\end{abstract}

%\date{\today}

\maketitle

%\tableofcontents

\section{\label{sec:level1}Skyrmions for Spintronics\protect\\}

Spintronics represents a new era in information storage, unburdened by the classical limits of today's computers. Operating based on electron spin, rather than charge, spintronic devices promise to overcome the scalability, energy consumption, and latency issues associated with current semiconductor electronics \cite{finocchio2021promise}, dramatically enhancing the efficiency of electronic devices while enriching them with new functionalities. Candidate quantum systems typically fall into two regimes, concerned with either the control of single localized spins on atomic sites in crystals \cite{awschalom2013quantum}, or rather with spin transport and dynamics in macroscale systems \cite{pulizzi2012spintronics}. Together, these systems are driving rapid advancements in non-volatile high-density memories and fast low-power operations in logic devices \cite{bader2010spintronics,pulizzi2012spintronics,finocchio2021promise}.

\begin{figure}
\includegraphics[width = \columnwidth]{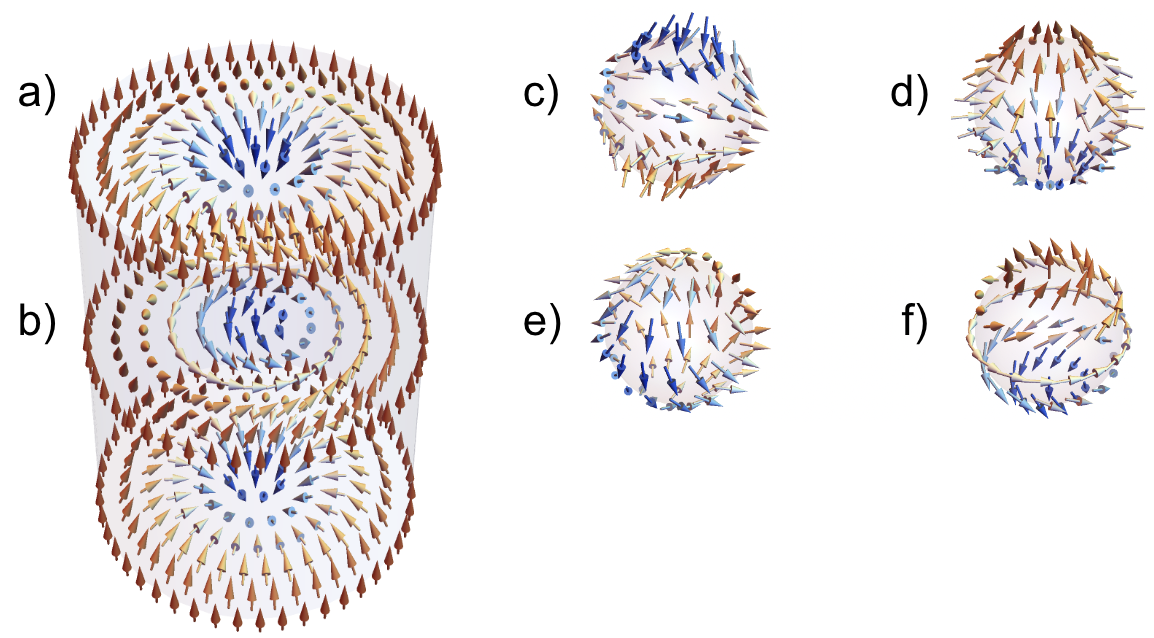}
\caption{Illustration of a magnetic skyrmion and it's typical spin configurations as a function of penetration depth. N\'eel-type skyrmions tend to stabilize at surfaces and in thin film systems (a), while Bloch skyrmions tend to form in bulk systems after some finite depth (b). Stereographic projections onto the unit sphere about the x-axis (left column) and z-axis (right column) are shown in plots c-f for both the N\'eel-type (c-d) and Bloch-type (e-f) skyrmions, respectively. Both spin textures have a topological charge of $|Q| = 1$ as highlighted by their complete mapping over the unit spheres. }
\label{fig:sk_tube}
\end{figure}

Historically, the engineering of quantum materials and architectures for spintronic devices has been geared towards systems of reduced dimensionality \cite{vzutic2004spintronics}. However, breakthroughs in characterization techniques of bulk quantum materials are reshaping modern approaches, unveiling a multitude of exciting three-dimensional magnetic systems which offer new symmetries, degrees of freedom, and dynamical behaviors/controls which reimagine current spintronic frameworks. 
Skyrmions are one of the most advanced building materials for spintronic devices, motivating a new stream of spintronics, coined skyrmionics, which relies on the couplings between two quantum degrees of freedom in real-space---spin and topology \cite{fert2017magnetic,finocchio2016magnetic}. 
Magnetic skyrmions define a unique class of topological objects, characterized by a multi-directional twisting of spins, typically induced by chiral interactions in bulk crystals and thin-films with special kinds of atomic symmetries \cite{nagaosa2013topological}. These magnetic modulations generate swirling vortex-like spin configurations on nanometric lengthscales, which can be regarded as knots in the magnetization (see Fig.~\ref{fig:sk_tube}) \cite{nagaosa2013topological}. This non-trivial winding (or knotting) in spin, defines a property known as the topological charge, $Q$, which quantifies the number of integer times the spin directions cover the surface of a sphere \cite{nagaosa2013topological}. This feature endows skyrmions with a sense of protection and robustness against unwinding perturbations, as singularities must be introduced to twist (or unknot) these spin textures, thereby posing an energy barrier to transformations between magnetic configurations with different $Q$, such as a uniformly magnetized state. The topology of the skyrmion can also be associated with a real-space Berry phase \cite{everschor2014real}, which manifests a myriad of emergent dynamics and novel transport phenomena, such as Topological and Skyrmion Hall Effects \cite{neubauer2009topological,jiang2017direct}, ultra-low current driven motion \cite{jonietz2010spin,yu2012skyrmion}, and multiferroic behavior \cite{seki2012observation,okamura2013microwave}. As a result skyrmions behave as rigid emergent quasi-particles, existing in a localized finite spatial extent with a quantized topological charge, capable of controlled motion, creation, and annihilation, and exhibiting emergent electrodynamics that transform Maxwell’s equations to a symmetric form in emergent space. This collection of properties makes skyrmions ideal candidates for information carriers in future information storage and logic technologies \cite{zhang2020skyrmion,kang2016skyrmion,fert2017magnetic,yu2017room,zhang2015magnetic}. %above room temp??

While skyrmions are commonly visualized as planar sheets in two-dimensions, in three-dimensions they extend into strings \cite{roessler2006spontaneous}. These strings may exist as isolated excitations \cite{seki2022direct} or alternatively condense into crystalline orders to form lattice arrangements. The configurational degrees of freedom and dimensionality of skyrmions are governed by a combination of magnetic interaction terms, sample geometry, and surface-to-volume effects. While skyrmions in thin and confined systems are dictated by surface and boundary effects, skyrmions in uninhibited bulk systems are dictated by volume effects which govern attractive interactions in conical phases and repulsive interactions in saturated phases \cite{loudon2018direct,leonov2023mechanism}. These two regimes lead to fundamental differences in skyrmion stabilization, defect formation, and interaction mechanisms. Note that here, our quantification of thin systems corresponds to sample thicknesses on the order of the period of the skyrmion spin spiral. In these systems, skyrmions may exist only in two-dimensions as surface states, or extend a finite amount, forming deformed structures such as chiral bobbers \cite{zheng2018experimental,rybakov2015new}. Similarly, confined systems are those in which there is only space to accommodate a handful of skyrmions, given their lattice period. As a result, skyrmions in thin/confined systems tend to be dominated by magnetization surface twists, exhibiting distortions such as chiral twists \cite{meynell2014surface,leonov2016chiral}, axial tube modulations \cite{wolf2022unveiling}, and edge states \cite{crisanti2020position,du2015edge, birch2020real} which fundamentally alter energetics of the system, giving rise to phenomena such as preferential nucleation at edges and boundaries, polar ordering, and surface-related defects \cite{du2015edge,han2020scaling,yu2020real,leonov2022skyrmion}. Conversely, skyrmions in bulk systems are stabilized through the three-dimensional proliferation of defects \cite{birch2021topological,milde2013unwinding}, forming a zoology of stable cores, lattices, clusters, superstructures, and transitions thereof which are mediated by skyrmion-skyrmion interactions and their tendency to reduce their energy \cite{leonov2022skyrmion}. 
These properties lead to extended complex three-dimensional filamentary and localized hybrid magnetization textures, such as magnetic torons, hopfions, skyrmion braids, and skyrmion chains (see Table \ref{table:1} for brief overview of various selected topological structures)  \cite{zheng2021magnetic,muller2020coupled,leonov2018homogeneous,leonov2022skyrmion,sallermann2023stability}. Such states may be stabilized naturally in bulk systems due to coupling to defects and distortions in the conical ground state \cite{muller2020coupled}, whereas thin systems may require geometric confinement or engineered defects in the vicinity of
opposite surfaces of the film \cite{leonov2018homogeneous}.  

\begin{table*}
\centering
\setlength{\tabcolsep}{5pt} % Adjusted to make more room for content
\renewcommand{\arraystretch}{1.5} % Increased line spacing for better readability
\begin{tabularx}{\textwidth}{||c c c >{\raggedright\arraybackslash}X||} 
 \hline
 Topological Object & Topological Charge & Surface/Bulk & Select Properties \\ [0.5ex] 
 \hline\hline
 Skyrmion & $\pm n$ & Both & Breathing modes \cite{mochizuki2012spin}, rotational motion \cite{seki2020propagation}, dilation \cite{liu2020three}. \\ 
 Toron & $\pm 1$ & Both & Conversion between skyrmions and hopfions \cite{li2022mutual,bo2021spin,leonov2021surface}, microwave magnetic excitation modes \cite{li2022mutual}. \\
 Chiral Bobber & $\pm 1$ & Surface & Tunable hall angle \cite{gong2021current}, high-frequency resonant modes \cite{bassirian2022breathing}. \\
 Hopfion & $\pm n$ & Both & No skyrmion hall effect \cite{wang2019current,gobel2020topological}, forms composite objects with skyrmions for controlled transport along tube channels \cite{zheng2023hopfion}. \\
 Skyrmionium & 0 & Both & No skyrmion hall effect \cite{kolesnikov2018skyrmionium}, spin wave driven motion exceeding that of skyrmions \cite{shen2018motion}. \\ [1ex] 
 \hline
\end{tabularx}
\caption{Description of various three-dimensional topological objects and some of their selected characteristics. The winding number `n' represents an integer quantity, which for the special case of the hopfions, quantifies the hopf index, `H' \cite{kent2021creation}. Note that this list is not complete or exhaustive and that these objects may co-exist with skyrmions/each other, enabling multi-bit encoding schemes.}
\label{table:1}
\end{table*}

\section{\label{sec:level1}Characterizing Topological Spin-Textures in Three Dimensions\protect\\}

A lack of bulk probes for skyrmions has lead to a significant bias towards the study of thin systems, yielding a more rigorous characterization and understanding of their stabilization and dynamics. As a result, practical efforts devoted to the implementation of skyrmions in spintronic devices have focused mainly on thin systems \cite{sampaio2013nucleation}. Thin systems offer the advantage of precise control at the single-skyrmion level, providing opportunities for the in situ selective engineering of materials \cite{buttner2018theory}. The less explored collective creation mechanisms for skyrmions in bulk systems can therefore appear more daunting, from a control standpoint. However, this should not deter efforts as the enhanced dimensionality of bulk systems opens the door to new stabilization pathways, configurational degrees of freedom, three-dimensional torque fields, and dynamical modes like gapless motion \cite{leonov2022skyrmion,muller2017magnetic}, rotation, string propagation, and dilation \cite{okamura2013microwave,onose2012observation,yokouchi2018current}, which offer unique functionalities and  logic device infrastructures to those of thin systems. In particular, bulk systems possess near-room temperature skyrmion phases with enhanced stability through quenching processes and defect pinning \cite{karube2020metastable,bannenberg2019multiple,henderson2022skyrmion,birch2019increased}, transport through string spin excitations \cite{yokouchi2018current}, electrical nucleation and motion at substantially lower current densities than those for thin films \cite{yu2017current}, and a diverse set of defects which facilitate restructuring and transition dynamics \cite{birch2021topological,gilbert2019precipitating,henderson2022skyrmion}. In spite of this, three-dimensional skyrmions in bulk systems remain widely unexplored.  

\begin{figure*}
\includegraphics[width = 150mm]{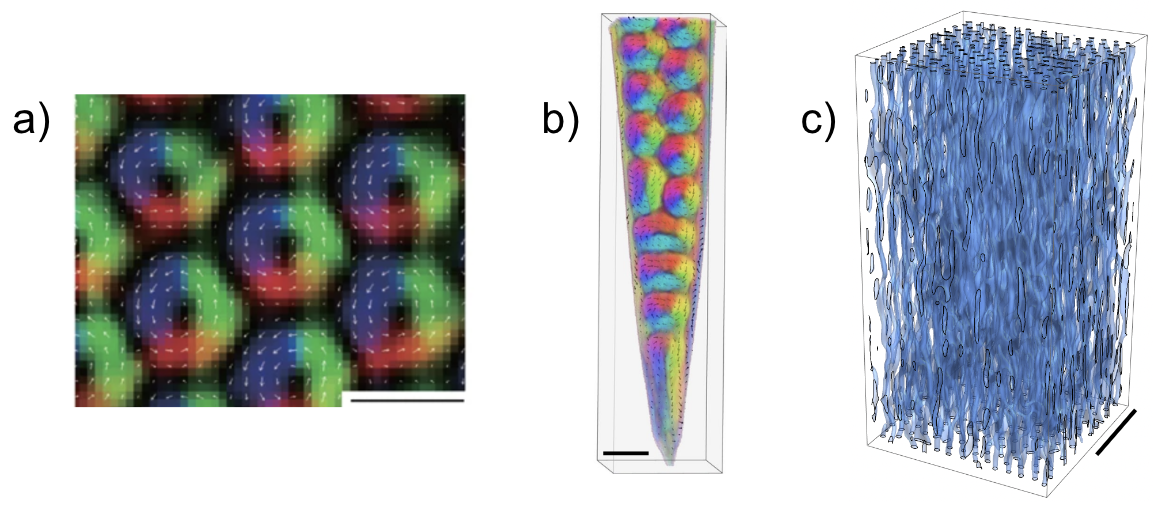}
\caption{Outline of skyrmion sample characterizations across three different probing techniques, highlighting various sample geometries and their respective imaging volumes. (a) An imaging slice containing the in-plane magnetization is shown for a skyrmion lattice in a thin film sample of Fe$_{0.5}$Co$_{0.5}$Si using Lorentz Transmission Electron Microscopy, reproduced from \cite{yu2010real}. (b) The magnetic induction of a confined skyrmion lattice using Vector-Field Electron Tomography is shown in three dimensions for a needle-shaped sample in FeGe, reproduced from \cite{wolf2022unveiling}. (c) The three-dimensional spin configuration of a bulk skyrmion lattice is shown for a millimeter-sized cube using Small Angle Neutron Scattering tomography, reproduced from \cite{henderson2023three}. Scale bars for plots (a-c) correspond to lengthscales of 90 nm, 100 nm, and 1 $\mu$m, with skyrmion lattice periods spanning 90 nm, 70 nm, and 130 nm, respectively.} 
\label{fig:sk_probes}
\end{figure*}

The most popular methods for skyrmion characterization can be categorized into either direct real-space imaging techniques, such as microscopy and holography, or indirect reciprocal space techniques, such as small angle neutron scattering (SANS), which require prior knowledge of the system for a complete structural characterization. Both of these investigative techniques suffer serious limitations; the former is only able to study confined and thinned samples due to the limited penetration depths of electrons and x-rays which are used as the probing particles (Fig.~\ref{fig:sk_probes}) \cite{yu2022real,seki2022direct,wolf2022unveiling}. The latter technique only provides 2D averaged datasets, incapable of distinguishing topologically distinct states which manifest the same magnetic scattering vectors \cite{white2018direct,kanazawa2012possible}. Therefore, three-dimensional experimental inquiries of bulk skyrmions have thus far remained elusive. 

Henderson et al. \cite{henderson2023three}, have developed a novel neutron scattering tomography technique which provides three-dimensional visualization of bulk micromagnetic materials using 3D angular datasets of 2D SANS images \cite{henderson2023three}. This technique was based on \cite{heacock2020neutron} and was applied to study the thermal equilibrium skyrmion triangular lattice phase of a bulk site-disordered material, Co$_{8}$Zn$_{8}$Mn$_{4}$. The results provided for the time three-dimensional visualizations of a bulk skyrmion lattice, uncovering exotic topological structures and defect-mediated transition pathways \cite{henderson2023three}. This technique unveils previously inaccessible skyrmion dynamics/features, providing insight into novel skyrmion stabilization pathways (such as through topological charge conserving magnetic toron elongation \cite{henderson2022skyrmion}), interaction and survival mechanisms (such as through twisting and memory effects), and ordering processes, serving as a novel route to understanding skyrmions \cite{henderson2023three}. In the past, bulk systems have been overlooked as skyrmion hosting materials for future devices owing to the seemingly random and chaotic nature of disorder and defects which was thought to complicate their control and inhibit their motion. However, the combination of these three-dimensional results and disordered skyrmion manipulations demonstrated in \cite{henderson2022skyrmion} completely dismantle those notions, demonstrating disorder in bulk systems as an additional tunable control parameter which enhances stability and drives new collective dynamics through exotic structures and pathways, thereby increasing the skyrmion phase space of the material. The result of this first application has already demonstrated that stabilization and transport properties/phenomena are not well understood for extended skyrmion structures, in spite of these structures holding great promise as future information carriers. This new quantum measurement technique therefore demands the re-innovation of our approaches to skyrmion implementations in spintronic devices now that we may have access to a comprehensive three-dimensional understanding of skyrmion nucleation, annihilation, transition, and organizational pathways across a wide parameter space, under diverse stimuli.       

Exploiting all of the available quantum degrees of freedom of both probing techniques and sample states is critical to the complete characterization and manipulation of quantum materials. In particular, neutrons can be manipulated to possess a quantum property called orbital angular momentum (OAM) which manifests helical wavefronts defined by a topological charge \cite{clark2015controlling,sarenac2019generation,sarenac2022experimental}. Utilizing this quantum property will enable new and direct methods of characterizing magnetic topological materials through tunable topological scattering modalities. A recent study presented in \cite{sarenac2022experimental} provides the first realization of single-valued OAM generation on nanometric lengthscales. Using microfabricated 2D arrays of fork dislocation phase-gratings in a conventional SANS setup, reference \cite{sarenac2022experimental} has revealed the doughnut shaped far-field intensity signature
of various topological neutron states. This technique effectively embeds the scattering interaction in a topological subspace, enabling unprecedented insights into topological features and excitations in quantum materials through topological interaction mechanisms such as targeted scattering, conservation of topological charge, and dynamic OAM transfer \cite{afanasev2019schwinger,afanasev2021elastic,gao2024dynamical,yang2018photonic,afanasev2018radiative}. Helical neutron beams may be used to directly probe topological features of bulk materials, such as correlation lengths, topological charge, and defect densities, thereby serving as a standalone measurement technique for topological phases, ushering in a new phase of quantum devices for quantum material characterization, which transcend the neutron scattering techniques discussed here. A simple construction of a few such experiments involves inserting an OAM neutron diffraction grating upstream of a skyrmion sample, followed by:
\begin{itemize}
  \item Examining the coherent interference between neutron and sample topology.
  \item Tuning the external magnetic field of the skyrmion sample between the helical and ferromagnetic phase boundaries, controllably changing the total topological charge/defect densities of the system, to study the topological dependence of OAM scattering signals.
  \item Rotating the sample-field configuration relative to the OAM grating to uncover the angular dependence of the OAM-topology scattering interaction.
  \item Applying this method in conjunction with SANS tomography to correlate defect densities with OAM scattering signatures
\end{itemize}
From these experiments, one might expect enhanced transmission of the OAM states in the convolved diffraction orders or enhanced scattering from the topological structures, while polarized implementations with spin-dependant OAM diffraction orders could induce dynamic rotations of the lattice through OAM-magnon modes. Furthermore, these methods may be extended to study additional topological features in both real and reciprocal space by incorporating helical wavefronts of various symmetries and topologies to input probe beams and their analyzers. For example, implementation of these methods across Angle-Resolved Photoemission Spectroscopy (ARPES) and neutron interferometry devices would enable the direct examination of topological interactions and excitations in quantum materials across a diverse range of lengthscales, dimensions, and degrees of freedom, spanning energy to spin. Together, these quantum measurement techniques reveal an unaddressed regime of physics in today's quantum materials, calling to question the fundamental approaches of current spintronic device methods.

\section{\label{sec:level1} Functionalities and Control of Three-Dimensional Topological Objects \protect\\}

Access to the topological character and third-dimension of skyrmion tube, enables the boundless study and discovery of new defect types, topological structures, and drive dynamics. While disorder has been revealed as a tunable parameter which facilitates skyrmion stability and dynamics through novel pathways, future investigations may lead to exotic conceptualizations involving higher topology skyrmion structures \cite{tang2021magnetic, rybakov2019chiral,zhang2016high,kolesnikov2018skyrmionium} and superstructures \cite{zheng2021magnetic,leonov2022skyrmion,foster2019two}, or entirely new three-dimensional topological quasiparticles as logical bits \cite{jiang2015blowing,kent2021creation,wu2022hopfions,zheng2018experimental,redies2019distinct,liu2020three}. For example, higher topology skyrmion bundles offer lower depinning threshold currents and faster propagation than traditional interfacial skyrmion structures \cite{tang2021magnetic}, while hopfions and topologically trivial skyrmioniums exhibit vanishing skyrmion hall effects which naturally support unidirectional current-driven motion for information transfer in devices \cite{liu2020three,tang2021magnetic}.  
Moreover, hybrid topological skyrmion structures such as torons \cite{henderson2023three,leonov2018homogeneous,li2022mutual}, chiral bobbers \cite{zheng2018experimental,rybakov2015new}, skyrmion braids \cite{zheng2021magnetic,zheng2023hopfion}, skyrmion chains \cite{du2015edge,leonov2022skyrmion}, and hopfion rings \cite{zheng2023hopfion} possess unique emergent features \cite{liu2020three,redies2019distinct,zheng2021magnetic,liu2022emergent,liu2020three,redies2019distinct} which gives rise to tunable dynamical modes \cite{hu2021unidirectional,bassirian2022breathing,liu2020three,bo2021spin,khodzhaev2022hopfion,gong2021current} and distinct transport signatures \cite{li2022mutual,gong2021current,bassirian2022breathing,zheng2021magnetic,ma2015microwave,liu2020three,raftrey2021field,liu2022emergent,gobel2020topological,tang2021magnetic,redies2019distinct,gong2021current,gobel2020topological}. These phenomena present novel pathways for the manipulation and identification of topological objects in future devices which both simplify their collective control and enhance their individual detection. No longer reliant on the compositional and geometrical engineering methods which severely inhibit existing device architectures \cite{fert2013skyrmions,muller2017magnetic,zhu2021current,shigenaga2023harnessing,shibata2013towards,zhang2023magnetic,zhang2020skyrmion,iwasaki2013current}, three-dimensional topological objects instead offer local dynamic methods of control based on tunable external stimuli which effectively serve as the ``knobs" of the device.  

Bit encoding schemes in three-dimensional skyrmion device architectures are markedly enriched as compared to their modern two-dimensional counterparts. In particular, traditional racetrack architectures rely on fixed position distances to encode bits, yielding storage and computing obstacles associated with thermal drift, reduced skyrmion stability, and low packing densities \cite{zheng2018experimental,schutte2014inertia}. Three-dimensional devices circumvent these issues, offering novel encoding opportunities through physical observables which are only realizable and controllable in three-dimensions, such as topology, helicity, string length, and the sequences of topological objects themselves. 
Routes for controlling these objects and their topology may derive from both traditional and contemporary methods involving applications of electric currents \cite{jonietz2010spin,yu2012skyrmion} and variations of magnetic fields \cite{tang2021magnetic}, respectively. While sufficient for topology-based encoding schemes, helicity-based encoding schemes pose additional in-situ control challenges.

Robust encoding schemes require the local dynamical control of the skyrmion helicity across the full solid-state device, rather than the global static control associated with engineered crystal chiralities \cite{shibata2013towards}. Recent developments involving electric field tuning in frustrated systems have established novel mechanisms for the dynamic control of skyrmion helicity \cite{yao2020controlling}, however, implementation of helical control protocols across a diverse set of magnetic systems demands a more universal means which does not directly rely on centrosymmetric crystal structures. These alternative methods may derive from external excitations involving photons, spin waves, and electric field pulses---excluding magnetic field manipulations and their associated device size, speed, and energy consumption challenges. In the former two control stimuli, OAM wave modes may be used to transfer their topologies to the magnetic systems, inducing chiral flips and nucleating objects with OAM-dependent helicities and topologies \cite{porfirev2023light,fujita2017encoding,fujita2017ultrafast}. These methods also offer additional dynamic controls for logic operations and information transfer involving object tweezing, rotation, and magnonic waveguides \cite{jiang2020twisted,yang2018photonic,xing2020magnetic}. Further helicity control using electric fields may be achieved through spin-orbit coupling methods in heterostructures \cite{ba2021electric,fillion2022gate}, or may be made possible through magnetoelectric interactions similar to those shown in magnetic vorticies \cite{yu2020nondestructive}. Physical implementations of such electric field control schemes may involve depth-mediated gate voltage protocols in ferromagnet/oxide interfaces in magnetic multilayers through fractional skyrmion tubes and modification of the interfacial Dzyaloshinskii–Moriya interaction (DMI) \cite{fillion2022gate,chen2023encoding}.

While future devices may not necessarily be constructed from millimeter-sized bulk samples like the one probed in \cite{henderson2023three}, the measurement techniques described in \cite{henderson2023three} and \cite{sarenac2022experimental} may be applied to the holistic study of diverse three-dimensional topological magnetic features and their novel transport phenomena, which can be employed in future spintronic logic devices. For example, skyrmion Field Effect Transistor's (FET) represent a promising avenue for low-power logic devices, operating on spin-polarized current driven motion \cite{kang2016skyrmion}. However, current two-dimensional skyrmion approaches have remained unrealizable owing to the Skyrmion Hall effect, where a transverse bending of the skyrmion motion relative to the drive current can cause unwanted annihilation events on device faces/edges \cite{hong2019magnetic}. Exploiting the third dimension of the skyrmion tube, new FET schemes can be developed which redefine bits as topological objects with vanishing skyrmion hall effects, such as hopfions, or introduce perpendicular skyrmion bit architectures which rely on the confined one-dimensional motion of perpendicular skyrmion tubes along channels in the helical background. In particular, nonaxisymmetric skyrmions, where the tubes form perpendicular to the external field, enable complex skyrmion cluster and chain formations which may provide new confined channels under which skyrmions, and other bits like merons, may move rapidly for logic operations \cite{muller2017magnetic,leonov2018toggle,sohn2019real}. This device architecture would not only enable the controlled directional transport of three-dimensional topological structures, but would realize the multi-bit encoding schemes with drastically enhanced packing densities as discussed above \cite{leonov2022skyrmion}. 
\begin{figure}[hbt!]
\includegraphics[width = \columnwidth]{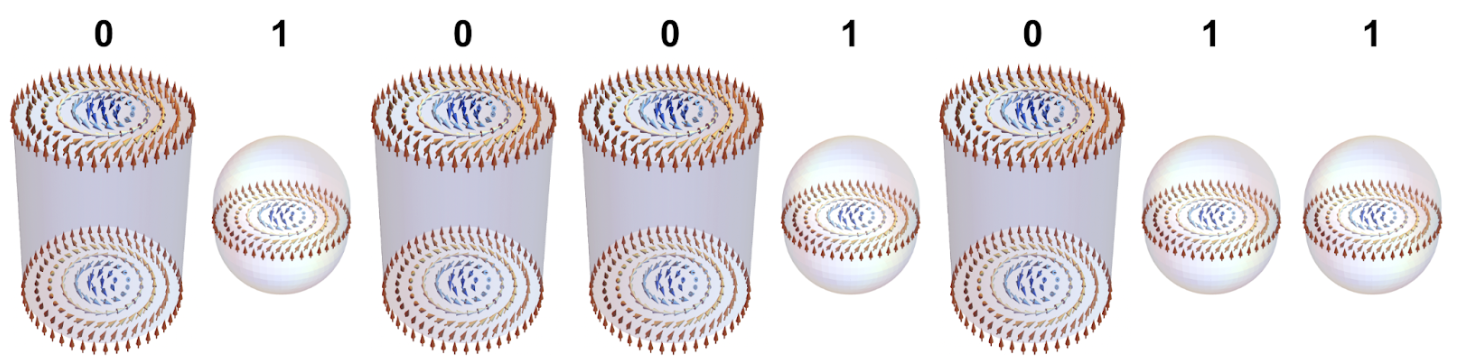}
\caption{Illustration of a three-dimensional topology-based bit encoding scheme using combinations of skyrmions (0) and torons (1). Encoding bits as distinct topological structures, rather than quantized distances between skyrmions, circumvents existing challenges associated with thermal drift, bit stability, and packing densities. Unprecedented control and readout of the bits can be achieved through their distinct dynamical modes and excitation signatures.}   
\label{fig:sk_device}
\end{figure}

\begin{figure*}
\includegraphics[width = 150mm]{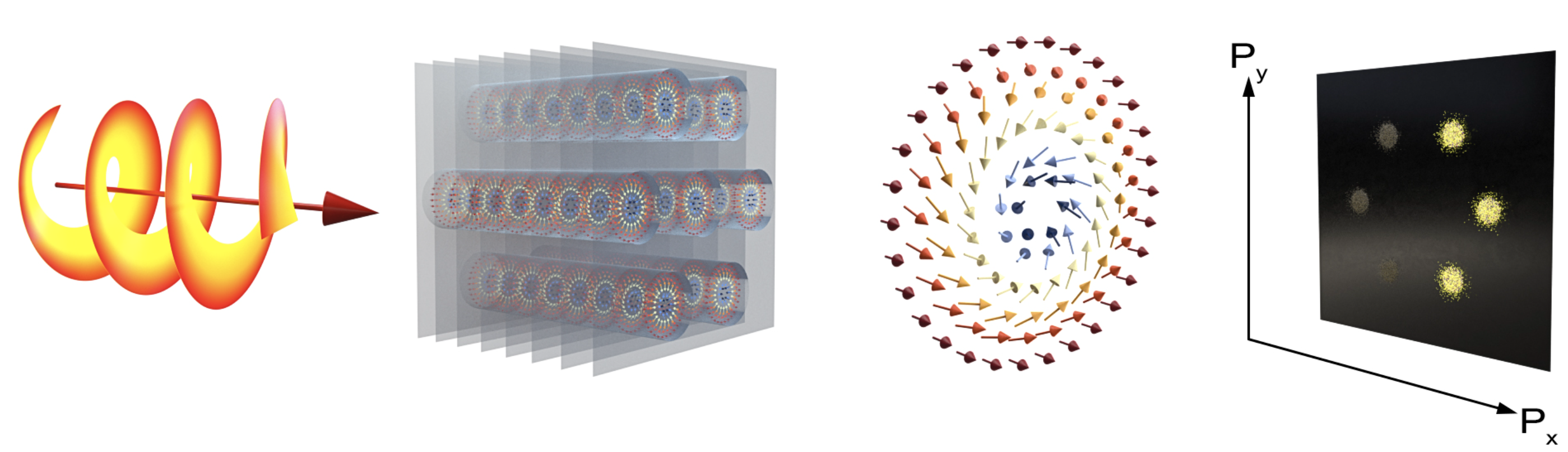}
\caption{Schematic of an ``operando'' SANS experiment involving the coherent control and measurement of skyrmion helicity as a function of depth. Input neutron OAM states on the left travel towards the skyrmion sample, with rotating helical wavefronts  propagating along the red arrow. These states are then scattered from the skyrmion sample which represents a heterostructure where distinct layers are defined by grey planes. 2D magnetization slices are shown for each plane, illustrating various possible skyrmion helicites through control of the interfacial DMI. The output states from the sample represent spin-OAM coupled states, where the topology of the skyrmion spin was encoded into the scattered neutrons as illustrated by the 3D spin vectors of the neutron beam after the sample. These states produce six-fold hexagonal diffraction patterns which are then analyzed in the transverse $P_{x}P_{y}$ polarization plane. The skyrmion helicity can be controlled at finite depths along the heterostructure sample, as highlighted by the grey planes, using electric field control methods outlined earlier. Using polarization analysis in the $P_{x}P_{y}$ plane \cite{sarenac2023novel}, the extinction direction of the scattering intensity of the peaks can be observed and used to directly determine the helicity. Incorporating tomographic rotations about the vertical axis, enables depth-resolved helicity determinations.}   
\label{fig:sans_exp}
\end{figure*}
 
Additional examples of three-dimensional device architectures may draw from composite topological structures and sequences such as hopfions rings or combinations of skyrmions, torons, skyrmioniums, and chiral bobbers. Using hopfion rings as bits would enable encoding through both the skyrmion and Hopf topological charge, three-dimensional directionally controlled transport/interactions along skyrmion tube channels, and detection via distinct hall signatures \cite{gobel2020topological}. Similarly, devices employing topological bit sequences would provide additional encoding degrees of freedom and dynamical modes for control and detection. More specifically, logic and readout designs for such three-dimensional encoding schemes may employ novel methods of signal multiplexing, topological conversion, and unidirectional information transfer through dynamical processes such as topological binning \cite{kind2021magnetic,chen2020skyrmionic}, field-induced transitions, and reprogrammable confined nanochannels \cite{wagner2016magnetic,seki2020propagation}. To achieve such functionalities and operations, these schemes would draw from collective bit-specific topological phenomena, encompassing exotic hall effects, magnetoresistive effects, and resonant spin wave dynamics discussed earlier. Physical device implementations could range from a simple homogeneous slice of a skyrmion sample to stacks of repeated layer structures, forming tailored multilayers. In the former, topology-based encoding schemes could be achieved using combinations of skyrmions and torons; in the latter, depth dependant skyrmion helicities could be achieved through interface engineering and selective dynamical controls (see Fig.~\ref{fig:sk_device}).

\section{\label{sec:level1}Dynamic Manipulations of Topological Quantum Numbers in SANS \protect\\}

Combining the quantum access afforded by structured neutron scattering methods, with the novel control schemes of three-dimensional topological objects, establishes a new route for the dynamic creation and control of future magnetic qubits which reimagines spintronic devices. Preliminary dynamic topological and helical controls of a skyrmion qubit could be demonstrated through an ``operando'' SANS experiment involving the coherent control of skyrmion helicity in three-dimensional heterostructure samples using structured neutron beams with coupled spin and OAM degrees of freedom. Exploiting three-dimensional material engineering to generate skyrmion structures with depth-mediated helicities, in combination with three-dimensional topological neutron scattering methods to selectively coupled to spin-orbit modes, would provide unprecedented control and enhanced visibility for magnetic structures. Here, one could toggle the skyrmion helicity in-situ between N\'eel- and Bloch-type skyrmions as a function of depth, while scattering tunable spin-orbit neutron beams as the sample is rotated (see Fig.~\ref{fig:sans_exp}). In doing so, one could dynamically control the helical quantum degree of freedom and selectively readout both the helical and topological quantum degrees of freedom through direct topological interference, OAM multiplexing, and spin-orbit tomography methods. This would serve as the first direct, real-time visualization and manipulation of three-dimensional topological and helical degrees of freedom, paving the way for future devices.

Ultimately, the neutron scattering quantum measurement probes discussed here have opened the door to an entirely new world of three-dimensional skyrmion physics and spintronics which overturns previous architectures of skyrmion nucleation and control, reshaping future quantum devices. These new devices possess computing infrastructures and tunable knobs which present drastic enhancements to existing spintronic technologies, overcoming fundamental challenges associated with stability, directional transport, size, and energy consumption. Together, these neutron methods address an unexplored regime of quantum scattering, uncovering three-dimensional topological phenomena and dynamics which are pioneering a new generation of spintronic devices with potentially superior computing architectures and dynamic controls.

\begin{acknowledgments}
This work was supported by the Canadian Excellence Research Chairs (CERC) program, the Natural Sciences and Engineering Council of Canada (NSERC) Discovery program, the Canada First Research Excellence Fund (CFREF), and the U.S. Department of Energy, Office of Nuclear Physics, under Interagency Agreement 89243019SSC000025. A portion of this research used resources at the High Flux Isotope Reactor, a DOE Office of Science User Facility operated by the Oak Ridge National Laboratory.

\end{acknowledgments}

\section{\label{sec:level3}Bibliography\protect\\}
\bibliography{refs}% Produces the bibliography via BibTeX.

\end{document}